\documentclass[twocolumn,aps,pra]{revtex4}
\usepackage{mathrsfs, mathbbol}
\usepackage{amsmath}
\usepackage{amssymb}
\usepackage{graphicx}
\usepackage{eufrak}
\usepackage{cancel}
 \usepackage{verbatim}

\newcommand{\ket}[1]{\left| #1 \: \right\rangle}
\newcommand{\bra}[1]{\left\langle \: #1  \right|}
\newcommand{\out}[2]{{\ket{#1}\hspace{-0.1cm}\bra{#2}}}

\newcommand{\expecfull}[3]{\bra{#1} \: #2 \:\ket{#3}}
\newcommand{\expec}[1]{\langle \: #1 \:\rangle}
\newcommand{\bigexpec}[1]{\left \langle #1 \right\rangle}
\newcommand{\lr}[1]{ \left( #1 \right)}

\newcommand{\bea}{\begin{eqnarray}}
\newcommand{\eea}{\end{eqnarray}}
\newcommand{\Tr}[1]{{\rm Tr} \left( #1 \right)}

\newcommand{\com}[2]{ \left[ #1,#2 \right]}
\newcommand{\acom}[2]{ \{ #1,#2 \}}
\newcommand{\lra}[1]{ \left| #1 \right|}

\newcommand{\lrb}[1]{ \left[ #1 \right]}

\newcommand{\sumlim}[2]{\sum\limits_{#1}^{#2}}

\newcommand{\so}[2]{\mathfrak{so}\lr{#1,#2}}
\newcommand{\su}[2]{\mathfrak{su}\lr{#1,#2}}
\newcommand{\binomial}[2]{ \left( \begin{array}{c}
#1\\#2
\end{array} \right)}

\def \r{\rho}
\def \k{\kappa}
\def \a{\alpha}
\def \b{\beta}
\def \d{\delta}

\def \o{\omega}

\def \ah{\hat{a}}
\def \ahd{\hat{a}^{\dagger}}
\def \bh{\hat{b}}
\def \bhd{\hat{b}^{\dagger}}
\def \ch{\hat{c}}
\def \chd{\hat{c}^{\dagger}}

\def \nn{\nonumber}
\def \nnl{\nonumber\\}

\begin{document}

\title{Computable negativity in two mode squeezing subject to dissipation}

\author{Marcin Dukalski and  Yaroslav M. Blanter%
}
\affiliation{Kavli Institute of Nanoscience, Delft University of Technology, Lorentzweg 1, 2628 CJ Delft, The Netherlands}

\begin{abstract} We study a system of two bosonic fields subject to two-mode squeezing in the presence of dissipation. We find the Lie algebra governing the dynamics of the problem and use the Wei-Norman method to determine the solutions.
Using this scheme we arrive at a closed form expression for an infinitely dimensional density operator which we use to calculate the degree of entanglement (quantified by Horodeckis' negativity) between the modes. We compare our result  to the known continuous variable entanglement measures. We analyse the conditions for entanglement generation and the influence of thermal environments on the state formed. The problem is relevant, in particular, for understanding of quantum dynamics of coupled optical and/or mechanical modes in optomechanical and nanomechanical systems.
\end{abstract}
\pacs{  }
\maketitle

\section{Introduction} \label{introduction}

Entanglement is a fascinating, unique, and classically unparallelled feature of quantum mechanics. In optical systems entanglement is produced by means of nonlinear media through a process of three- or more- wave mixing, spontaneous down-conversion or two mode squeezing \cite{ScullyBook}, where in each case dissipation and the temperature of the environment play a negligible role. Modern quantum optomechanical \cite{Aspelmeyer2010,Marquardt} and nanomechanical \cite{LabadzeDukalski2014} systems equipped with elements with sufficiently strong nonlinearities  could give rise to similar effective squeezing phenomena. Such bosonic systems are subject to  dissipation  through coupling to the classical environment, which might reduce the degree of produced entanglement.
It is therefore very important  to get a quantitative understanding of the squeezing versus  dissipation interplay, with the aim of producing or maintaining entanglement between bosonic degrees of freedom such as photons (light or microwave quanta) or phonons (vibrational quanta).

The main obstacle in determining the environmentally induced effects on the degree of bosonic entanglement formed is twofold. For one, the Hilbert space of both of the bosonic modes is infinitely dimensional with high $n$ Fock states $\ket{n}$ contributing to the degree of entanglement formed, and this requires the usage of more advanced entanglement measures. Secondly, the presence of dissipation requires
solving, instead of a Schr\"{o}dinger equation,  a more complicated Lindblad type master equation which governs the time evolution of the open quantum system \cite{Lindblad1976,Kossakowski1972}.
The two conditions combined also mean that the solutions to the equations of motion should be determined for an arbitrarily large Hilbert space, requiring us to find the complete infinite set of time dependent density matrix elements $\r_{ij,kl}$. The key to determining the solutions, and hence finding the degree of entanglement, will be to recast the problem using the Wei-Norman method (WNM) \cite{WeiNorman63,WeiNorman64}, something that we have already accomplished  in a similar problem of dissipative one mode squeezing in one of our previous works \cite{Dukalski2014SingleMode}.
The WNM amounts to translating  from the master equation containing  non-commutating (super-)operators  to a set of new nonlinear differential equations, whose degree of nonlinearity grows with the complexity of the commutators structure.

In the past the Wei-Norman method has been applied to study Lie algebra valued problems \cite{ban92,Gerry85,Gerry87,Dattoli86,Dattoli87,Dattoli88}, many of which found applications in optical systems. These problems, however, mainly involved Lie algebras spanned by no more than three generators, and if any extensions to the algebras were considered, only their reductions were studied  \cite{Twamley93,cervero1996generalized}. Here we will treat the most general two-mode squeezed problem subject to dissipation involving as many as fifteen generators, and we will be able to obtain analytical solutions to a reduced problem spanned by a Lie algebra composed of ten,  and in a special case all fifteen  generators.

The great advantage of the Wei-Norman method is that it will provide a density operator in a closed form, which we could use to evaluate the generally established entanglement measures, such as negativity \cite{Horodecki1996,Vidal2002,Horodecki07}. We will be able to find an analytical form of negativity stemming  from the  operator structure alone, and as a result of it one can use it to find the amount of entanglement in the system in a generalised problem involving time dependent squeeze parameters, non-Markovian baths, or different environments of individual modes, as long as the bosonic operator form remains unchanged. Furthermore, we will show that the explicitly calculated negativity of an infinitely dimensional density operator in some cases is completely compatible with the separability criterion of the continuous variable states \cite{Simon,LuMingDuan} for this system. Explicit calculation of negativity though, not only allows one to find whether or not the state is separable, but also how large is the degree of entanglement.
Moreover, due to a similarity in the Lie algebra structures, compared to the single mode squeezing case, we will see how the solutions show different quantitative and qualitative behaviour  in  two distinct system parameter regimes (like these found in \cite{Dukalski2014SingleMode}), and how in a two-mode symmetric system the entanglement measure is the same in both regimes.

This work is structured as follows. In Section \ref{system}, upon introducing  the two mode squeezing Hamiltonian, we will couple it to a Markovian bath and transfer the Lindblad type superoperator  master equation  into a Lie algebra valued problem, where superoperators present in the equations are identified with Lie algebra elements. Further, we present the generic form of the solution of the problem with the system initialised in a vacuum state, which allows us in Section \ref{entanglement measures} to determine the  degree of entanglement present in the system.
In Section \ref{CVSSC}, we compare this result and steaming from it separability condition to that obtained from continuous variable separability condition. Afterwards, in Section \ref{Implications} we present the explicit solutions to the master equation and interpret the separability condition in terms of the bath temperature, proving that regardless squeezing strength and dissipation rate the state is inseparable at zero temperature. Additionally we also investigate the effects  of the asymmetric  modes-baths coupling strengths and deviation from resonance between the two squeezed modes  and the driving mode.
Finally, in Section \ref{thermal} we  study the case of the system initially in the thermal equilibrium with the environment and we point out a finite time interval of separability in the transient regime and further show that the same conditions as before are obeyed in the steady state regime. We close with the conclusions.

\section{System} \label{system}
Let us consider a system comprised of   three coupled bosonic modes,
with their annihilation operators $\ah, \bh,\hat c$, where $\hat c$ denotes a strong driving mode, and $\hat a$ and $\hat b$ are the output modes, with their respective frequencies $\o_c$, $\o_a$ and $\o_b$. In the presence of a dominant $x^3$-type coupling, upon a modal decomposition, and in the regime $\o_c\sim \o_a+\o_b$, the cross interaction terms of the type $\chd\ahd\bhd$, $\chd\ahd\bh$  or  $\chd\ah\bhd$,  will be suppressed in the rotating wave approximation \footnote{Alternatively these can be considered to be energy violating terms, which are negligible outside of the ultra-strong coupling regime}. In this case the system Hamiltonian will take the form
\bea
\hat{H}=\o_a \ahd\ah+\o_b\bhd\bh+\o_c\chd\ch+g\lr{\chd\ah\bh+\ch\ahd\bhd}\,,\nn
\eea
which in the interaction picture with respect to all modes takes the form
\bea
\hat{\mathcal{V}}&=&g\lr{\chd\ah\bh e^{i\delta t}+\ch\ahd\bhd e^{-i\delta t}}\nnl
&=&g c\lr{\ah\bh e^{i\lr{\delta t+\eta}}+ \ahd\bhd e^{-i\lr{\delta t+\eta}}}\,,\nn
\eea
where ${\delta}=\o_c-\o_b-\o_a$, and where in the second step we have assumed that the driving mode $\ch$ is a very large coherent state with a real amplitude $c$ and phase $\eta$, and thus will be considered to be \emph{frozen out}.
We can further remove the explicit time dependence of this Hamiltonian by yet another unitary transformation $U^\dagger \hat{H}U-iU^\dagger\partial_t U $, where $U=\exp\lrb{i{\delta} t\lr{ \ahd\ah+ \bhd\bh}}$ and redefining $g c e^{-i \eta}\to\frac{\xi}{4}$ we get
\begin{align*}
\hat{H}={\delta}\lr{\ahd\ah+ \bhd\bh}+\frac{\xi}{4}\ahd\bhd+\frac{\xi^*}{4}\ah\bh\,,
\end{align*}
with the factor of one-quarter included in the definition  for future convenience. Assuming a standard coupling of every mode to their respective environments, e.g. phonon or a photon continuum, and assuming that the interaction kernel in each case is time local (Markovian approximation), we arrive at a Lindblad type master equation
\begin{align}
\dot\r&=-i\com{\hat{H}}{\r}+\k_1 \lr{n_{1,th}+1}\mathcal{D}_{\ah}\r+\k_{1} n_{1,th}\mathcal{D}_{\ahd}\r\nnl
&~~~~~~~~~ +\k_2\lr{n_{2,th}+1}\mathcal{D}_{\bh}\r+\k_2 n_{2,th}\mathcal{D}_{\bhd}\r\,,\label{masterEq}
\end{align}
where $\mathcal{D}_{\Theta}\r=\Theta\r\Theta^{\dagger}-\frac{1}{2}\acom{\Theta^{\dagger}\Theta}{\r}$, $\k_i$ is the dissipation rate and  $n_{i,th}=\lr{e^{\o_j/k_b T_i}-1}^{-1}$ is the thermal occupation number  in the bath at temperature $T_i$ of bosonic mode $i=1,2$ (given by operators $\ah$ and $\bh$ respectively).\\

For the system initially in the vacuum state $\ket{00}$, the equation (\ref{masterEq}) has a solution,  that can be written  in a form
{\small \bea
\r\lr{t}&=&\mathcal{N}\exp\lrb{f_3  H_3+f_5  H_5+f_9  H_9+f_{12}  H_{12}}\out{00}{00}\,,\label{solution}\\
\mathcal{N}&=&\lr{1-x_+}\lr{1-x_-}, ~~~x_{\pm}=f_3\pm\sqrt{f_5^2+f_9^2+f_{12}^2}\,,\nn
\eea}
where the numerical prefactor marks the trace-normalisation condition and where
{\small \begin{align*}
 H_{3}\r  &=\ahd \r  \ah+\bhd \r  \bh\,,  &  H_{9}\r  &=\ahd \r  \ah-\bhd \r  \bh\,,\\
 H_{5}\r  &=i e^{-i\varphi}\ahd\bhd  \r -i e^{i\varphi}\r   \ah\bh\,, & 
 H_{12}\r &=e^{-i\varphi}\ahd\bhd \r +e^{i\varphi}\r  \ah\bh\,,
\end{align*}}
are four of the fifteen  elements of the $\so{4}{2}$ Lie algebra presented and elaborated on in the Appendix; the remaining eleven generators drop out due to the initial condition choice. Here $\varphi=\pi-{\rm Arg}\lr{\xi}$. The solution (\ref{solution}), despite being written in a compact form, spans the whole of the infinite dimensional Hilbert space of both modes.
Moreover, the exponents of the superoperators should be understood either in terms of Taylor expansions or in terms of  matrix exponents of the matrices operating on the product space
\begin{align}
\rho=\hat{A}\out{n}{m}\hat{B}~~\to~~\tilde{\rho}=\hat{A}\otimes\hat{B}^{\dagger}\ket{n}\otimes\ket{m}\,.
\end{align}
In this work we will adapt the Taylor series approach. 

Lastly, quite remarkably thanks to this construction one can obtain analytical expressions for  moments
$$
\expec{\ah^{\dagger k}\ah^l\bh^{\dagger m}\bh^n}=\Tr{\ah^{\dagger k}\ah^l\bh^{\dagger m}\bh^n\r\lr{t}}\,,
$$
in terms of functions $f_i$ by skilfully differentiating with respect to $f_3,f_5,f_9$ and/or $f_{12}$ and then renormalising the moment generating function
\bea
\Lambda &=&\Tr{\exp\lrb{f_3  H_3+f_5  H_5+f_9  H_9+f_{12}  H_{12}}}=\mathcal{N}^{-1}\,,\nn
\eea
for example
\bea
\expec{\ah^{\dagger  }\ah }=\frac{1}{2}\mathcal{N}\lr{\partial_{f_3}+\partial_{f_9}} \mathcal{N}^{-1}\,,
\eea
where  from the solution we can see that the only non-zero moments must be of the form $\expec{\ah^{x+y-z}\ah^{\dagger z}\bh^{y}\bh^{\dagger z}}$  with $x,y,z\in\mathbb{Z}$ and $x+y\geq z$. This automatically implies that the joined power of the moment must be even, and that of the simplest (quadratic) moments the only non-zero ones are $\expec{\ahd\ah}$, $\expec{\bhd\bh}$, $\expec{\ah\bh}$ and $\expec{\ahd\bhd}$. We will need these in the next section we will study the entanglement stored in this bosonic system.

  The time- and system parameters-dependent \emph{real} functions $f_3, f_5 ,f_9$ and $f_{12}$  are determined using the Wei-Norman method and obey the complicated set of first order nonlinear differential equations presented in the Appendix. We postpone the discussion about how the solutions are obtained to Section V and  first, in Sections III and IV, we focus on the entanglement measures as the discussion in terms of the functions $f_i$ is more transparent.

\section{Entanglement measures} \label{entanglement measures}
The solutions in equation (\ref{solution}) are described by the application of exponents of creation  super-operators on a two-mode vacuum state. If we wish to work with an exact solution and \emph{not} truncate the Taylor expansion of the super-operator exponent we arrive at an infinitely dimensional  density operator $\r=\out{\psi}{\psi}$ of a potentially entangled state $\ket{\psi}$. In such a case the finite dimensional entanglement measures no longer apply, which is a reason why here we will attempt to use negativity  \cite{Horodecki1996,Vidal2002,Horodecki07} which is not limited by the dimensional restrictions\footnote{Other than the presence of bound entanglement in systems with dimensions greater that $2\times 3$.
As we will see in the sections to come,  bound entanglement is unlikely to be present.}. The result (\ref{solution}) could also be interpreted as a continuous variable state (CVS), where we could use the entanglement measure bounds imposed by the conditions first presented in Ref. \cite{Simon,LuMingDuan}. Here we will show that we can calculate the negativity explicitly, which we will later compare to the CVS separability criterion \cite{Simon, LuMingDuan}. Both of these measures in their core rely on the partial-transposition $pT$ operation, 
 given by
\bea
\lr{\out{ij}{kl}}^{pT_1}=\out{kj}{il}~~~~\lr{\out{ij}{kl}}^{pT_2}=\out{il}{kj}\,,\nn
\eea
i.e. the transposition is taken only with respect to the first and second subspace respectively, and we define the negativity as a sum of negative eigenvalues of $\rho^{pT}$. 
For entangled states defined in a ${\rm dim}\times {\rm dim}$ dimensional Hilbert space, negativity is  a monotonously growing function (an entanglement monotone) with a range  $\lrb{0,\frac{1}{2}\lr{{\rm dim}-1}}$, giving zero for   separable states.

Determination of negativity is not an easy task however, due to the dimensionality dependence of this entanglement monotone and the infinitely dimensional density operator. Here every application of $H_{3},H_{5}, H_{9}, H_{12}$ gives rise to a   yet larger Hilbert space and the exponent of these operators results in an infinitely long Taylor expansion. Moreover, negativity is based around negative eigenvalues, which need to be determined. Here we outline the sketch of a proof which is presented in detail in the Appendix.

Finding the eigenvalues in this setting amounts to finding the roots of a characteristic polynomial of infinite degree, i.e.  to solving the equation  $\det\lrb{\r^{pTr} -I \lambda}=0$.
 Using the determinant of a matrix exponent -- exponent of a trace relation we can write
\bea
\det\lrb{\r^{\rm pTr} -I \lambda}=\det\lrb{-\lambda I}\exp\lr{ \sumlim{j=1}{\infty}-\frac{\Tr{\lr{\r^{\rm pTr}}^j}}{j\lambda^j}}\,,\nn
\eea
which, thanks to the  property of the form of the solution (\ref{solution}),
\begin{align*}
\Tr{\lr{\r^{\rm pTr}}^j}=\frac{\lr{1-x_+}^j\lr{1-x_-}^j}{\lr{1-x_+^j}\lr{1-x_-^j}}\,,
\end{align*}
yields
\bea
\det\lrb{\r^{\rm pTr} -I \lambda}={\det\lrb{-\lambda I}} \prod\limits_{p,q=0}^{\infty}\lr{1-\frac{x_+^{p} x_-^{q}\mathcal{N}}{\lambda}}\,.\nn
\eea
where $x_{\pm}$ were defined before, and where the eigenvalues can be directly read out.  Since (as we will later show) $f_3$ is always be positive, the only negative eigenvalues will have the form $x_+^{p} x_-^{2q+1}$ provided that $x_-<0$. Upon adding all of them up we obtain  the negativity
\begin{align}
{\rm Neg}&=\lra{\mathcal{N}\sum_{p,q=0}^{\infty} x_+^{p} x_-^{2q+1}}=\frac{x_-}{1+x_-}\nnl
&={\rm Max}\lr{0,\frac{-f_3+\sqrt{f_5^2+f_9^2+f_{12}^2}}{1+f_3-\sqrt{f_5^2+f_9^2+f_{12}^2}}}\,.\label{eq:negativity}
\end{align}
This is the main result of this paper.

\section{Continuous Variable States Separability Condition} \label{CVSSC}
As first simultaneously and independently formulated by \cite{Simon,LuMingDuan} the continuous variable states separability criterion stems from quadratic   relations of the type
{\small \begin{align}
\bigexpec{\lr{\Delta \hat{X}_{\vec d}}^2}+\bigexpec{\lr{\Delta \hat{X}_{\vec d'}}^2}\geq \lra{d_1 d_2'-d_2 d_1'+d_3 d_4'-d_4 d_3'}\,,\label{firstcond}
\end{align}}
which is the Heisenberg uncertainty relation obeyed by all states, with $d_i$ and $d'_i$ being components of the real $\vec d$ and $\vec d'$ four-vectors,  $\hat{X}_{\vec v}=v_1 \hat{x}_1+v_2 \hat{p}_1+v_3 \hat{x}_2+v_4 \hat{p}_2$, and $\Delta \hat{A}=\hat{A}-\expec{\hat A}$.
Separable states, on the other hand, need to obey a more stricter inequality
{\small
\begin{align}
{\lr{\Delta \hat{X}_{\vec d}}^2}+\bigexpec{\lr{\Delta \hat{X}_{\vec d'}}^2}\geq&& \lra{d_1 d_2'-d_2 d_1'} +\lra{d_3 d_4'-d_4 d_3'}\,, \label{sepcond}
\end{align}
such that for the right combination of $\vec d$ and $\vec d'$ with $\lra{d_i}=\lra{d_i'}=1 ~\forall i$, the first relation is bounded from below by zero, and the second one can be bounded by four.

The uncertainty on the left hand side can be expressed as
\begin{align*}
\bigexpec{\lr{\Delta \hat{X}_{\vec d}}}&=\lra{\lambda_1}^2\lr{2\expec{\ahd\ah}+1}+\lra{\lambda_2}^2\lr{2\expec{\bhd\bh}+1}\nnl
&~~~~+4{\rm Re}\lr{\lambda_1\lambda_2\expec{\ahd\bhd}}\,,
\end{align*}
where $\lambda_1=\lr{d_1+i d_2}/\sqrt{2}$ and $\lambda_2=\lr{d_3+i d_4}/\sqrt{2}$, and where the other quadratic terms evaluate to zero for the state given by the equation (\ref{solution}).

By imposing that $\lra{\lambda_i}=\lra{\lambda_i'}=1$ we can find the optimal criterion for separability.  Next, without a loss of generality we can set ${\rm Arg}\lr{\lambda_1}=0$, impose the saturation of the lowest possible bound of the Heisenberg uncertainty principle by setting  $d_1 d_2'-d_2 d_1'+d_3 d_4'-d_4 d_3'=0$, and maximize $\lra{d_1 d_2'-d_2 d_1'}+\lra{d_3 d_4'-d_4 d_3'}=4$, which requires ${\rm Arg}\lr{\lambda_1'}={\rm Arg}\lr{\lambda_2}-{\rm Arg}\lr{\lambda_2'}=\frac{\pi}{2}$. This way we turn the separability condition (\ref{sepcond}) into
\begin{align*}
4\leq \frac{4\lr{1-f_3^2+\lra{z}^2+2 \lra{z}\cos\lr{{\rm Arg}\lr{\lambda_2}-{\rm Arg}\lr{z}}}}{\lr{1-\lr{f_3+\lra{z}}}\lr{1-\lr{f_3-\lra{z}}}}\,,
\end{align*}
where $z=e^{-i\varphi}\lr{f_{12}+i f_5}$ and where we have assumed identical baths $\k_1=\k_2$ and $n_{{\rm th}, 1}=n_{{\rm th}, 2}$ implying $f_9=0$, see discussion in the Appendix. This criterion has one left degree of freedom ${\rm Arg}\lr{\lambda_2}-{\rm Arg}\lr{z}$, which when fixed to be equal to $\frac{3\pi}{2}$ gives $f_3>\lra{z}$
-- the same separability criterion as that obtained from the explicit negativity calculation.
%

In the next sections we will use this result in combination with the solutions to the equations of motion to determine the system parameters separability condition.

\section{Implications of the separability conditions} \label{Implications}
In order to understand the separability condition in terms of the system parameters we need to first translate the master equation (\ref{masterEq}) into a set of equations for functions $f_i\lr{t}$ with the initial condition $f_i\lr{0}=0\forall i$. The details of the procedure are outlined in the Appendix, where we outline how an entire set of fifteen functions can be obtained. In this work so far we have focused on systems initialised in vacuum state $\r\lr{0}=\out{00}{00}$, allowing us to narrow our interest to but four functions, which independent of the initial condition $\r\lr{0}$, obey the following   set of equations
\begin{align}
 \dot{f}_{3}&= -\frac{1}{2}c_{21,+} f_{3} -\frac{1}{2}c_{21,-} f_{9} +c_{11,-}f_{3} f_{9}  -\frac{1}{2}\lra{\xi}  f_{3} f_{5}\nnl
&~~~~+\frac{1}{2} c_{11,+}\left(f_{3}^2+f_{5}^2+f_{9}^2+f_{12}^2\right)+c_{10,+} \,,\label{EQ3}\\
\dot{f}_{5}&= -\frac{1}{2} c_{21,+}f_{5} +c_{11,+}f_{3} f_{5} +c_{11,-}f_{5} f_{9} +2 \delta  f_{12}\nnl
&~~~~+\frac{1}{4}\lra{\xi}   \left(-f_{3}^2-f_{5}^2+f_{9}^2+f_{12}^2+1\right)\,, \label{EQ5}\\
\dot{f}_{9}&= -\frac{1}{2}c_{21,-} f_{3} -\frac{1}{2}c_{21,1+} f_{9} +c_{11,+}f_{3} f_{9} -\frac{1}{2}\lra{\xi}  f_{5} f_{9} \nnl
&~~~~+\frac{1}{2}c_{11,-}
   \left(f_{3}^2-f_{5}^2+f_{9}^2-f_{12}^2\right)+\frac{1}{2}c_{10,-}\,, \label{EQ9}\\
\dot{f}_{12}&= -\frac{1}{2} c_{21,+}f_{12} +c_{11,+}f_{3} f_{12} +c_{11,-}f_{9} f_{12} -2 \delta  f_{5}\nnl
&~~~~-\frac{1}{2}\lra{\xi}  f_{5}f_{12} \,,  \label{EQ12}
  \end{align}
where we have  defined
\begin{align*}
c_{xy,\pm}=\k_1 (x n_{{\rm th},1}+y)\pm \k_2 (x n_{{\rm th},2}+y)\,.
\end{align*}

It is easy to see that for an identical baths case  $\k_1=\k_2=\k$ and $n_{{\rm th}, 1}=n_{{\rm th}, 2}$, all $c_{xy,-}=0$, implying that $f_9\lr{t}=0$, and completely independently  in the resonant regime  $\d=0$ we have $f_{12}\lr{t}=0$. Moreover in the absence of dissipation, all
$c_{xy,\pm}=0$ both $f_{3}\lr{t}$ and $f_{9}\lr{t}$ vanish, and for $\xi=0$ we get $f_{5}\lr{t}=f_{12}\lr{t}=0$. All of these are examples of parameter and Lie algebra reductions leading to significant simplifications in the equations above, to the extent that the non-linear set of equations above can be solved analytically in a resonant identical baths case, and we were able to analytically determine the steady state solutions if either the baths are identical, or the system is driven resonantly, or both with $n_{{\rm th}, 1}=n_{{\rm th}, 2}=0$.

The solutions to the equations above fall into two parameter regimes with a baths populations independent boundary
\begin{align*}
{\Xi}^2=\lr{1- \frac{\lr{\k_1-\k_2}^2}{\lr{\k_1+\k_2}^2}}\lr{\d^2+\lr{\k_1+\k_2}^2}\,.
\end{align*}
 As a result we define the  underdamped $\lr{\lra{\xi}^2\geq\Xi^2}$, and overdamped $\lr{\lra{\xi}^2<\Xi^2}$ regime, which we call this way due to either unbounded or bounded expectation values $\expec{\ahd\ah}$ and $\expec{\bhd\bh}$  respectively.
 
One can verify numerically that both in the over- and the underdamped regime the equations (\ref{EQ3})-(\ref{EQ12}) possess steady state solutions. By setting the left-hand sides of these four equations to zero one can obtain the steady state values of $f_{3}$, $f_{5}$, $f_{9}$  and $f_{12}$ algebraically. The process yields a set of solutions larger than those obtained by considering the set of nonlinear ordinary  differential equations with initial conditions $f_i\lr{0}=0$, therefore the algebraic solutions found have been verified by solving the differential equations  numerically. The steady state solutions to the equations in the  overdamped regime results at zero bath temperature take the compact form
\begin{align*}
 f_3&= \frac{\lra{\xi} ^2 \left(\k_1^2+\k_2^2\right)}{8 \k_1 \k_2 \left(\delta ^2+(\k_1+\k_2)^2\right)-2 \k_1 \k_2 \lra{\xi}^2} \,,\\
 f_5&= \frac{2 \lra{\xi}  (\k_1+\k_2)}{4 \left(\delta ^2+(\k_1+\k_2)^2\right)-\lra{\xi} ^2}\,, \\
 f_9&= \frac{\lra{\xi} ^2 \left(\k_2^2-\k_1^2\right)}{8 \k_1 \k_2 \left(\delta ^2+(\k_1+\k_2)^2\right)-2 \k_1 \k_2 \lra{\xi}^2} \,,\\
 f_{12}&= -\frac{2 \delta  \lra{\xi} }{4 \left(\delta ^2+(\k_1+\k_2)^2\right)-\lra{\xi} ^2} \,,
\end{align*}
and the solutions in the $\lra{\xi}^2>\Xi^2$ regime are too incomprehensible to present here, which is why  we will also present the parameter simplified ones. 
%
The solutions in the detuned regime with identical non-zero temperature baths read
\begin{align*}
f_3&= \frac{2 \kappa  (2 n_{\rm th}+1)}{\sqrt{\lra{\xi} ^2-\delta ^2}+4 \kappa  (n_{\rm th}+1)}\,, \\
 \lra{z}&= \frac{\sqrt{\lra{\xi} ^2-\delta ^2}+2 \kappa }{\sqrt{\lra{\xi} ^2-\delta ^2}+4 \kappa  (n_{\rm th}+1)} \,,
\end{align*}
in the underdamped regime, and in the overdamped they become
\begin{align*}
f_3&= \frac{\lra{\xi} ^2+4 n_{\rm th} (n_{\rm th}+1) \left(\delta ^2+4 \kappa ^2\right)}{4 (n_{\rm th}+1)^2 \left(\delta ^2+4 \kappa ^2\right)-\lra{\xi} ^2}\,, \\
\lra{z}&= \frac{2\lra{\xi} \sqrt{\delta ^2+4 \kappa ^2} (2 n_{\rm th}  +1 )}{4 (n_{\rm th}+1)^2 \left(\delta ^2+4 \kappa ^2\right)-\lra{\xi} ^2}\,.
\end{align*}
The two sets of solutions above imply  that the separability condition
$f_3>\lra{z}$
reduces to
\begin{align}
4 n_{\rm th} \k > \sqrt{\lra{\xi}^2-\delta^2} &~~~~{\rm for}~~~~ \lra{\xi}^2\geq 4\k^2+\delta^2\label{cond1}\,,\\
2 n_{\rm th} \sqrt{4\k^2+\d^2} > \lra{\xi}    &~~~~{\rm for}~~~~ \lra{\xi}^2< 4\k^2+\delta^2\label{cond2}\,,
\end{align}
 where the parameter regime discontinuity in this result  is gone in the absence of detuning, and the same form is obeyed in both the under- and the over-damped regime, where negativity is described by a single function independent of the parameter regime. 
 
 Moreover, in this symmetric resonant regime, where one only needs to consider the solutions to the equations of motion for functions $f_{1-6}$, the other ones returning $f_{7-15}\lr{t}=0$, one can solve the complete set of differential equations analytically also in the transient regime. This has to do with the fact that the master equation is described by a set of operators spanning the $\so{2}{2}$ Lie algebra, which decomposes into two sets of $\su{1}{1}$ Lie algebras, with the set of six equations decoupling into two sets of three equations which independently can be solved by the method of quadratures. Identical separability conditions and the same expression for negativity
{\small
\begin{align}
{\rm Neg}={\rm Max}\lr{
\frac{2\left(1-e^{-t (\k +\lra{\xi}/2 )}\right) \left(\lra{\xi} -4 \k  n_{th}\right)}{  e^{-t (\k +\lra{\xi}/2 )} \left(\lra{\xi} -4 \k  n_{th}\right)+2\k  \left(2 n_{th}+1\right)}\;,0}\,,\label{entanglementf3f5}
\end{align}}
in either under- and overdamped regimes is reflected by the fact that only one of the  copies of the $\su{1}{1}$ Lie algebras determines the entanglement. Equation (\ref{entanglementf3f5}) shows that at $t=0$ the state is completely separable, i.e. ${\rm Neg}=0$, while in the steady state it inseparable provided that $\lra{\xi}>4 \k   n_{th}$, which is a parameter reduced expression (\ref{cond1}) and (\ref{cond2}). Moreover, at zero bath temperature i.e. $n_{\rm th}=0$ states violate the separability condition in all parameter regimes.
 Lastly, it is worth observing that in the absence of dissipation $f_5= \tanh \lra{\xi} t/4$ and all other $f_i=0$, which not only violates the separability condition for any $t>0$ and gives rise to a divergent negativity as $t\to \infty$.
 
\section{System initially in a thermal state} \label{thermal}
Since the temperature of the bath plays an important role in the separability condition, it is worth investigating the effect of the initial state's temperature on the steady state entanglement obtained.
%
%
In the previous section we have assumed that the system is initiated in the vacuum state $\ket{00}$, however in the presence of the environment at a non-zero temperature, this might be difficult to accomplish, and the state prior to two-mode squeeze driving should be initiated in a separable state $\r\lr{0}=\r_{\ah,th}\otimes\r_{\bh,th}$, where denote the thermal state density operators $\r_{\hat{c},th}=\exp\lrb{\hbar\o_{\hat{c}}\hat{c}^{\dagger}\hat{c}/k_bT_{\hat c}}/\Tr{\exp\lrb{\hbar\o_{\hat{c}}\hat{c}^{\dagger}\hat{c}/k_bT_{\hat c}}}$. Here we will treat the simplest case   of the initial condition already in equilibrium with the  environment such that $T_{\hat a}=T_{\hat b}$,  $\o_{\hat{a}}= \o_{\hat{b}}=\o$ and hence $n_{th,1}=n_{th,2}=\tau\lr{1-\tau}^{-1}$, with $\tau=e^{-\b\hbar \o}$.  The two modes still remain orthogonal, i.e. $\ah\neq \bh$. 
The initial condition now can be written as $\r\lr{0}=\r_{\ah,th}\otimes\r_{\bh,th}=\lr{1-\tau}^2e^{\tau H_3}\out{00}{00}$. With the suitably chosen normal ordered solution Ansatz
\begin{align*}
\r\lr{t}&=e^{f_0\lr{t}} e^{f_3\lr{t}H_3}e^{f_5\lr{t}H_5}e^{f_{1}\lr{t}H_{1}}e^{f_6\lr{t}H_6}\\
&\hspace{4cm}\times e^{f_2\lr{t}H_2}e^{f_4\lr{t}H_4}\r\lr{0}\,,
\end{align*}
where operators $H_{1,2,4,6}$ contain normal ordered annihilation operators. We see that   since   the system is no longer initiated in the vacuum, we cannot disregard a given set of exponents of operators.
Using  the form of the initial condition
\begin{align*}
\r\lr{t}&=\lr{1-\tau}^2e^{f_0\lr{t}} e^{f_3\lr{t}H_3}e^{f_5\lr{t}H_5}e^{f_{1}\lr{t}H_{1}}e^{f_6\lr{t}H_6}\\
&\hspace{3.2cm}e^{f_2\lr{t}H_2}e^{f_4\lr{t}H_4}e^{\tau H_3}\out{00}{00}\,,
\end{align*}
we can now   commute  $\exp\lrb{\tau H_3}$ through exponents of operators $H_{1,2,4,6}$, and re-decompose using the Wei-Norman scheme as presented in the Appendix. As a result  we obtain
{\small \bea
\r\lr{t}&=&\lrb{\lr{1- g_3}^2- g_5  ^2}e^{ g_3\lr{t} H_3}e^{g_5 {H}_{5}}\out{00}{00}\label{entanglementForm}\,,
\eea}
where $g_i\lr{t}=f_i\lr{t}+\mathcal{F}_i\lr{t}$, and where the only two relevant $\mathcal{F}_i$ as functions of $f_i$ read
\bea
\mathcal{F}_3\lr{t}&=&\frac{1}{2} \left(-\frac{e^{2 f_1+f_6} \tau }{f_2 \tau -f_4 \tau -1}-\frac{e^{2 f_1-f_6} \tau }{f_2 \tau +f_4 \tau -1}\right)\,,\nnl
\mathcal{F}_5\lr{t}&=&\frac{1}{2} \left(\frac{e^{2 f_1-f_6} \tau }{f_2 \tau +f_4 \tau -1}-\frac{e^{2 f_1+f_6} \tau }{f_2 \tau -f_4 \tau -1}\right)\,.\nn
\eea
The form of the later functions reflects how important it was to know the transients of all of the six functions $f_i$ as well as the apparent simplification in the problem allowed us to determine functions $\mathcal{F}_{3,5}$ to begin with.
By virtue of the form of equation (\ref{entanglementForm}), we can immediately state that negativity will take the form
\bea
{\rm Neg}
&=&{\rm Max}\lr{\frac{ \left(\lra{\xi} -4 \k  n_{th}\right)-e^{-t (\k +\lra{\xi}/2 )}\lra{\xi}  \left(2 n_{th}+1\right)}{2 \left(2 n_{th}+1\right) \left(\lra{\xi} e^{-t (\k +\lra{\xi} )} +2\k  \right)},0}\,.\nn
\eea
 Moreover, this result in comparison to the equation (\ref{entanglementf3f5}) has the same steady state amount of entanglement $\frac{\lra{\xi} -4\k n_{ th }}{4 \k \lr{2 n_{th}+1}}$, however the key difference is that when the state starts in a thermal equilibrium with the environment its negativity remains zero for a finite amount of time $t=2\lr{2\k +\lra{\xi} }^{-1}\log \left(\frac{\lra{\xi}  \lr{2n_{th}+1}/2 }{\lra{\xi} -4 \k n_{th}}\right)$, which only makes sense for the case of any entanglement formed, i.e. $\lra{\xi}>4 \k n_{\rm th}$ .

\section{Conclusions} \label{conclusions}
In this work we have shown that, one can use the Wei-Norman method to study analytically a bosonic entanglement process subject to dissipation. The Lie algebra valued description based solution Ansatz   allows one to calculate the exact expression of entanglement evolution or its steady state form as measured by negativity.   Additionally, we have shown that the negativity calculated from the solution is completely compatible with the continuous variable separability condition. Moreover, we have shown that for time independent system parameters, one can determine  analytically the solutions to the equations of motion in the Wei-Norman setting in the transient and the steady state. Finally, should the bipartite state be initially in thermal equilibrium with the environment, then the steady state entanglement does not change, however there is a finite amount of time in the transient regime where the degree of entanglement is lower compared to that when the state is  initialised in vacuum.

 The results formulated in this paper in terms of general functions $f_i$ remain applicable for (effective) two-mode driven systems with time dependent parameters (driving strength $\xi$, dissipation rates $\k_i$ or detuning $\delta$), which then require using the same equations with time dependent coefficients. Moreover, this method is very well suited for investigating similar problems of more than two modes with pairwise-squeezing interaction terms.
As a result such extensions can be very important in experiments investigating entanglement in continuous variable systems.

\section{Acknowledgements}

The authors wish to thank Giorgi Labadze and Antoni Borr\'{a}s for useful discussions.
This work was supported by the Netherlands Foundation for Fundamental Research on Matter (FOM).

\newpage
\section{appendix}
\subsection{The Lindblad equation of motion and the underlying Lie algebra structure}
The Lindblad type master equation (\ref{masterEq}) can be rewritten in the form
\bea
\dot\r &=&\sumlim{i=0}{15}\a_i H_i \r\,,\label{LieMasterEq}
\eea
where
{\small \begin{align*}
\a_{1}&=-\frac{1}{4}\lr{\k_1\lr{2n_{1,th}+1}+\k_2\lr{2n_{2,th}+1}},\\
\a_{2,8}&=\frac{1}{2}\lr{\k_1\lr{n_{1,th}+1}\pm\k_2\lr{n_{2,th}+1}},\\
\a_{7}&=-\frac{1}{2}\lr{\k_1\lr{2n_{1,th}+1}-\k_2\lr{2n_{2,th}+1}},\\
\a_{3,9}&=\frac{1}{2}\lr{\k_1 n_{1,th}\pm\k_2 n_{2,th}},~~~~~~\a_{0}=\frac{1}{2}\lr{\k_1+\k_2},\\
\a_{4}&=\a_{5}=\frac{1}{4}\lra{\xi},~~~~~~~~\a_{15}=2\d,
\end{align*} }
with all other $\a_i=0$,
and where
  {\small \begin{align*}
 H_{1,7}\r  &=\frac{1}{2} \lr{\ah \ahd \r +\ahd \ah \r +\r  \ah \ahd+\r  \ahd \ah}, \\
 &~~\pm\frac{1}{2}\lr{ \bh \bhd \r +\bhd \bh \r +\r  \bh \bhd+\r \bhd \bh} ,\\
 H_{2,8}\r  &=\ah \r  \ahd\pm\bh \r  \bhd,  \\
 H_{3,9}\r  &=\ahd \r  \ah\pm\bhd \r  \bh,  \\
 H_{4,11}\r  &=  e^{i\lr{\varphi+\lr{1\pm 1} \pi/4}}\ah\bh \r -e^{-i\lr{\varphi+\lr{1\pm 1} \pi/4}}\r  \ahd\bhd,   \\
 H_{5,12}\r  &= e^{-i\lr{\varphi-\lr{1\pm 1} \pi/4}}\ahd\bhd  \r -e^{i\lr{\varphi-\lr{1\pm 1} \pi/4}}\r   \ah\bh, \\
 H_{6,10}\r  &=\frac{1}{2} i \left(e^{-i\varphi}\bhd \r  \ahd-e^{i\varphi}\ah \r  \bh \mp e^{i\varphi}\bh \r  \ah \pm e^{-i\varphi}\ahd \r  \bhd\right), \\
 H_{13,14}\r &=\frac{1}{2} \left(e^{i\varphi}\ah \r  \bh \pm e^{i\varphi}\bh \r  \ah\pm e^{-i\varphi}\ahd \r  \bhd+e^{-i\varphi}\bhd \r  \ahd\right), \\
 H_{15}\r &=\frac{1}{2} i\lr{ \ahd \ah \r -\r  \ahd \ah+\bhd \bh \r -\r  \bhd \bh},
 \end{align*}}
with the first (second) index corresponding to the upper (lower) signs and with $H_0$ being the identity superoperator, i.e. $H_0 \r=\r$.

Thanks to the elementary commutation relation
\bea
\com{\Theta_i}{\Theta_j^{\dagger}}=\delta_{i,j}~~~{\rm and}~~~\com{\Theta_i}{\Theta_j}=0 \ ,
\eea
where $\Theta_1=\ah$ and $\Theta_2=\bh$, the set of fifteen superoperators closes under commutation (see Table \ref{tabso42}), thus forming a Lie algebra.
In what follows we define skew (anti-)symmetric matrices
\bea
L_{i,j}=E_{i,j}+E_{j,i}\,,~~~~~~K_{i,j}=E_{i,j}-E_{j,i}\,,\nn
\eea
where $E_{i,j}$ is a matrix with 1 in the $i^{\rm th}$ row and $j^{\rm th}$ column and zero elsewhere. It is easy to verify that the linear combinations of the above
  \bea
\begin{array}{rclrclrclrcl} 
 \mathcal{H}_{1}&=&-2 L_{2,3} & \mathcal{H}_{6}&=&L_{1,5}  &  \mathcal{H}_{7}&=&-L_{1,4}  \\
    \mathcal{H}_{10}&=&K_{4,5} & \mathcal{H}_{13}&=&L_{1,6}  & \mathcal{H}_{14}&=&K_{5,6} \\
     \mathcal{H}_{15}&=&K_{4,6}
 \end{array}\nn
 \eea
 and
 \bea
 \begin{array}{rclrcl}
  \mathcal{H}_{2,3}&=&K_{1,2}\mp L_{1,3}   & \mathcal{H}_{4,5}&=&L_{2,5}\pm K_{3,5}  \\
   \mathcal{H}_{8,9}&=&K_{3,4}\pm L_{2,4} &\mathcal{H}_{11,12}&=&-K_{3,6}\mp L_{2,6}
 \end{array}\nn
 \eea
 obey the same commutation relations, and that the   $L_{i,j}$ and $K_{i,j}$ above are elements of the $\mathfrak{so}\lr{4,2}$ Lie algebra. Upon a homomorphism $H_i\to\mathcal{H}_i$ we can show that the superoperators from the master equation (\ref{masterEq}) are just a different incarnation of the $\mathfrak{so}\lr{4,2}$ Lie algebra.

 Realising that, for time independent $\k_i$, $n_{th,i}$, $\delta$ and $\xi$, the solution to the master equation (\ref{masterEq}) in the form of (\ref{LieMasterEq}) is simply
 \bea
\r\lr{t} &=&\exp\lrb{t\sumlim{i=0}{15}\a_i H_i} \r\lr{0}\,,\label{LieMasterSol}
\eea
which is simply given by a Lie group element acting on the initial state. This can be thought of as a \emph{rotation}, or a movement on the   surface embedded in six dimensions satisfying the equation
 \bea
  1=-x_1^2-x_2^2+x_3^2+x_4^2+x_5^2+x_6^2\,,\nn
 \eea
 which can be understood as the trace-preservation condition of the density operator $\r$  \cite{salmistraro1993invariants}.
 The form of equation (\ref{LieMasterSol}) however  is not very useful for any purposes, and we will proceed with the so called Wei-Norman method \cite{WeiNorman63,WeiNorman64}, to decompose the right hand side of equation (\ref{LieMasterSol}), however this method is much more powerful and allows one to solve the equation (\ref{masterEq}) for time dependent $\k_i$, $n_{th,i}$, $\delta$ and $\xi$, allowing for studying modulated squeeze-driving and non-Markovian baths.



\begin{table*}[ht]
{\small
\begin{tabular}{c|ccccccccccccccc|}
$ \com{\cdot}{\cdot}  $       & $H_{1}$    & $H_{2}$   & $H_{3}$ & $H_{4}$ & $H_{5}$ &\multicolumn{1}{c|}{$H_{6}$} & $H_{7}$ & $H_{8}$ & $H_{9}$ & \multicolumn{1}{c|}{$H_{10}$} & $H_{11}$ & $H_{12}$ & $H_{13}$ &  $H_{14}$& \multicolumn{1}{|c|}{$H_{15}$}  \\\hline
 $H_{1}$  & 0 & -2 $H_{2}$& 2 $H_{3}$ & -2 $H_{4}$ & 2 $H_{5}$&\multicolumn{1}{c|}{0} & 0 & -2 $H_{8}$ & 2 $H_{9}$ & \multicolumn{1}{c|}{0} & -2 $H_{11}$ & 2 $H_{12}$ & 0 & \multicolumn{1}{|c}{0} & \multicolumn{1}{|c|}{0} \\
 $H_{2}$  & 2 $H_{2}$  & 0      & $H_{1}$ & 0 & 2 $H_{6}$ & \multicolumn{1}{c|}{$-H_{4}$} & $H_{8}$ & 0 & 2 $H_{7}$ & \multicolumn{1}{c|}{0} & 0 & 2 $H_{13}$ & $H_{11}$ & \multicolumn{1}{|c}{0} & \multicolumn{1}{|c|}{0} \\
 $H_{3}$  & -2 $H_{3}$ & $-H_{1}$  & 0 & 2 $H_{6}$ & 0 & \multicolumn{1}{c|}{$-H_{5}$} & $-H_{9}$ & -2 $H_{7}$ & 0 & \multicolumn{1}{c|}{0} & -2 $H_{13}$ & 0 & $-H_{12}$ & \multicolumn{1}{|c}{0} & \multicolumn{1}{|c|}{0} \\
 $H_{4}$  & 2 $H_{4}$  & 0      & -2 $H_{6}$ & 0 & $-H_{1}$ & \multicolumn{1}{c|}{$-H_{2}$} & 0 & 0 & 2 $H_{10}$ & \multicolumn{1}{c|}{$-H_{8}$} & 0 & 2 $H_{15}$ & 0 & \multicolumn{1}{|c}{0} & \multicolumn{1}{|c|}{$-H_{11}$} \\
 $H_{5}$  & -2 $H_{5}$ & -2 $H_{6}$& 0 & $H_{1}$ & 0 & \multicolumn{1}{c|}{$-H_{3}$} &    0 & -2 $H_{10}$ & 0 & \multicolumn{1}{c|}{$H_{9}$} & -2 $H_{15}$ & 0 & 0 & \multicolumn{1}{|c}{0} & \multicolumn{1}{|c|}{$H_{12}$} \\
 $H_{6}$  & 0   & $H_{4}$   & $H_{5}$ & $H_{2}$ & $H_{3}$ & \multicolumn{1}{c|}{0} & $H_{10}$ & 0 & 0 & \multicolumn{1}{c|}{$H_{7}$} & 0 & 0 & $H_{15}$ & \multicolumn{1}{|c}{0} & \multicolumn{1}{|c|}{$H_{13}$} \\\cline{1-7}\cline{12-14}\cline{16-16}
 $H_{7}$  & 0 & $-H_{8}$  & $H_{9}$ & 0 & 0 & $-H_{10}$ & 0 & $-H_{2}$ & $H_{3}$ & \multicolumn{1}{c|}{$-H_{6}$} & 0 & 0 & $-H_{14}$ & $-H_{13}$ & 0 \\
 $H_{8}$  & 2 $H_{8}$  & 0      & 2 $H_{7}$ & 0 & 2 $H_{10}$ & 0 & $H_{2}$ & 0 & $H_{1}$ &\multicolumn{1}{c|}{ $H_{4}$} & 0 & 2 $H_{14}$ & 0 & $-H_{11}$ & 0 \\
 $H_{9}$  & -2 $H_{9}$ & -2 $H_{7}$& 0 & -2 $H_{10}$ & 0 & 0 & $-H_{3}$ & $-H_{1}$ & 0 & \multicolumn{1}{c|}{$-H_{5}$} & 2 $H_{14}$ & 0 & 0 & $-H_{12}$ & 0 \\
 $H_{10}$ & 0       & 0      & 0 & $H_{8}$ & $-H_{9}$ & $-H_{7}$ & $H_{6}$ & $-H_{4}$ & $H_{5}$ & \multicolumn{1}{c|}{0} & 0 & 0 & 0 & $-H_{15}$ & $H_{14}$ \\\cline{1-14}\cline{16-16}
 $H_{11}$ & 2 $H_{11}$ & 0      & 2 $H_{13}$ & 0 & 2 $H_{15}$ & \multicolumn{1}{c|}{0} & 0 & 0 & -2 $H_{14}$ & \multicolumn{1}{c|}{0} & 0 & $H_{1}$ & $H_{2}$ & \multicolumn{1}{|c}{$H_{8}$} & \multicolumn{1}{|c|}{$H_{4}$} \\
 $H_{12}$ & -2 $H_{12}$&-2 $H_{13}$& 0 & -2 $H_{15}$ & 0 & \multicolumn{1}{c|}{0} & 0 & -2 $H_{14}$ & 0 & \multicolumn{1}{c|}{0} & $-H_{1}$ & 0 & $-H_{3}$ & \multicolumn{1}{|c}{$H_{9}$} & \multicolumn{1}{|c|}{$-H_{5}$} \\
 $H_{13}$ & 0       & $-H_{11}$ & $H_{12}$ & 0 & 0 & \multicolumn{1}{c|}{$-H_{15}$} & $H_{14}$ & 0 & 0 & \multicolumn{1}{c|}{0} & $-H_{2}$ & $H_{3}$ & 0 & \multicolumn{1}{|c}{$H_{7}$} & \multicolumn{1}{|c|}{$-H_{6}$} \\\cline{1-7}\cline{12-14}\cline{16-16}
  $H_{14}$ & 0       & 0      & 0 & 0 & 0 & 0 & $H_{13}$ & $H_{11}$ & $H_{12}$ & $H_{15}$ & $-H_{8}$ & $-H_{9}$ & $-H_{7}$ & 0 & $-H_{10}$\\\cline{1-7}\cline{12-14}\cline{16-16}
 $H_{15}$ & 0       & 0      & 0 & $H_{11}$ & $-H_{12}$ & \multicolumn{1}{c|}{$-H_{13}$} & 0 & 0 & 0 & \multicolumn{1}{c|}{$-H_{14}$} & $-H_{4}$ & $H_{5}$ & $H_{6}$ & \multicolumn{1}{|c}{$H_{10}$} & \multicolumn{1}{|c|}{0} \\\hline
\end{tabular}}
\caption{Commutation relation table. The smallest (middle) $6\times 6$ ($10\times 10$) box encloses the $\mathfrak{so}(2,2)$ ($\mathfrak{so}(3,2)$)  Lie algebra.}\label{tabso42}
\end{table*}

 \subsection{Wei-Norman method treatment and the resultant equations of motion}
We take the Ansatz
\begin{align*}
\r\lr{t}=e^{f_0\lr{t}} \prod\limits_{i} e^{f_i\lr{t}H_i} \r\lr{0}\,,\nn
\end{align*}
with the ordering $i=3,5,9,12,1,7,15,6,14,10,13,2,8,$ $4,11$, with the exponent of $H_3$ acting last and the exponent of  $H_{11}$ acting first on the initial condition $ \r\lr{0}$. From the definition of the super-operators $H_{i}$, one can see that the ordering chosen above is normal, i.e. annihilation (creation) super-operators acting first (last), and in the middle super-operators which are composed of creation and annihilation operators
\footnote{The ordering of the indices might not seem very natural, however here we had to make a choice between transparent Lie sub-algebra division (discussed later) and re-naming of the superoperators such that the ordering is $i=1,2,...,15$; in this work we chose for the former.}. Lastly, it is important to note that a different ordering Ansatz will result in a different set of equations for functions $f_i\lr{t}$.

The set of resultant differential equations is quite complicated and non-transparent, thus outside the scope of this article. In a special, very convenient, case when $\r\lr{0}=\out{00}{00}$, we see that the set of eleven right-most operators acting on the initial condition leaves it unchanged, with the exception of $e^{f_1H_1}$ which contributes an overall scaling factor, thus in combination with the only decoupled equation of motion $\dot{f}_0=\frac{1}{2}\lr{\k_1+\k_2}$, gives
\bea
\r\lr{t}&=&e^{\frac{1}{2}\lr{\k_1+\k_2}t+2f_1} e^{f_3\lr{t}H_3}  e^{f_5\lr{t}H_5}  e^{f_9\lr{t}H_9}\\
&&~~~~~~~~~\times e^{f_{12}\lr{t}H_{12}} \out{00}{00}\nnl
&\equiv& e^{\frac{1}{2}\lr{\k_1+\k_2}t+2f_1}\tilde{\r} \ ,\label{eq:apprho2}
\eea
at which point the order does not matter due to the mutual commutativity of the remaining operators, and the scalar prefactor plays the role of a normalisation condition. Regardless of the initial condition,  the kinetic equations for functions $ f_3, f_5, f_9$, and $f_{12}$ were given in the main body of the text and the necessary equation for $f_1$   takes the form

  \begin{align*}
\dot{f}_{1}&= \frac{1}{2}  c_{11,+}f_{3}+\frac{1}{2} c_{11,-}f_{9} -\frac{1}{2} \lra{\xi}  f_{5}-\frac{1}{4}c_{21,+}  \ . \\
  \end{align*}
Using the set of equations (\ref{EQ3}-\ref{EQ12}) and the equation above one can verify by differentiating both sides and remembering the initial condition $f_i\lr{t=0}=0\forall i$, that
\begin{align*}
 \frac{e^{2 f_1+\frac{1}{2} t (\k_1+\k_2)}}{(1-f_{3})^2-f_{5}^2-f_{9}^2-f_{12}^2}=1 \ ,
\end{align*}
which allows us to eliminate the scale factor in equation (\ref{eq:apprho2}) in favour of functional dependence on functions $f_3, f_5, f_9$, and $f_{12}$.
 One can verify the trace-preserving nature of the evolution by taking the trace of the equation and arriving at $\partial_t\Tr{\r}=0$. 

Additionally, the tabular display of the operators into sub-groups marks  the use of smaller Lie algebras equation decompositions, and  surface dimensional reduction, such that:
\begin{itemize}
\item[1] $\delta=0$  decouples operators $H_{11-15}$  from the algebra ($H_{15}$ vanishes directly from the equation of motion, and $H_{11-14}$ do not enter the dynamics due to their commutation relation properties - see Table \ref{tabso42}).
\item[2] $\k_1 =\k_2 =\k\neq 0$ and $n_{th,1}=n_{th,2}$, decouples operators $H_{7-9}$, and effectively the operator $H_{10}$, and the system reduces again to $\so{3}{2}\subset\so{4}{2}$.
\item [3]$\xi= 0$, decouples operators $H_{4,5}$ and effectively $H_{6,11-14}$. The relevant operators form an  $\so{2}{2}\subset\so{4}{2}$ Lie algebra which decomposes into two copies of $\su{1}{1}$ Lie algebras acting in separate subspaces of $\ah$ and $\bh$ bosons.
\item [4]$\k_1=\k_2=0$, decouples operators $H_{2,3,6}$ and  $H_{7-10}$. The remaining operators form again an  $\so{2}{2}\subset\so{4}{2}$ Lie algebra which decomposes into two sets of operators acting separately from the right or from the left of the density operator. This has to do with the fact that said evolution no longer needs to be described using a von-Neumann equation in superoperators, but rather a Schr\"{o}dinger equation described only by right or left acting operators separately.
\item[5] Moreover, conditions {\bf 1} and {\bf 2} combined, also lead to an $\so{2}{2}$ reduction, where the decomposition into  two copies of $\su{1}{1}$ Lie algebras is different, and it resembles the decomposition found in \cite{Dukalski2014SingleMode}.
\end{itemize}

The Lie-algebraic reduction described above has to do with the reduced dimensionality of the space embedding the surface, to which  one can deem  the evolution to be confined, such that for the $\so{3}{2}$ Lie algebra case we are dealing with a five-dimensional space with a surface given by the equation $  1=-x_1^2-x_2^2+x_3^2+x_4^2+x_5^2$ and in the $\so{2}{2}\sim \su{1}{1}\otimes\su{1}{1}\sim\so{2}{1}\otimes\so{2}{1}$ the evolution is confined to a product space of two hyperboloids embedded in three dimensions.
%

\subsection{Computing the negativity}

 The solutions to the equations of motion written using the Wei-Norman conditioned on both modes initially in the vacuum state are given in equation (\ref{solution}).
For the purposes of this proof we will rewrite this result using the property of mutual commutation of the above operators
\begin{align}
\r\lr{t}&=\mathcal{N}e^{\lr{f_3+f_9}\ahd\cdot\ah}e^{\lr{f_3-f_9}\bhd\cdot\bh}\nnl
&~~~~\times e^{\lr{ f_{12}+i f_5}e^{-i\varphi} \ahd\bhd\cdot}e^{\lr{f_{12}-i f_5}e^{i\varphi} \cdot\ah\bh}\out{00}{00}\label{resemble3}
\end{align}
and we will  rewrite the matrix in terms of a quadriple infinite sum with redefinitions
$g_{\pm}=f_3\pm f_9$ and  $z=\lr{ f_{12}+i f_5}e^{-i\varphi}$
\begin{align*}
\r&=\mathcal{N}\sumlim{ijkl=0}{\infty}\frac{g_{+}^{i}g_{-}^{j}z^{k}\bar{z}^l}{i!j!k!l!} \lr{\ahd}^{i+k}\lr{\bhd}^{j+k}\out{00}{00} {\ah}^{i+l}{\bh}^{j+l}\\
&=\mathcal{N}\sumlim{ijkl=0}{\infty}\frac{g_{+}^{i}g_{-}^{j}z^{k}\bar{z}^l}{i!j!k!l!} \sqrt{\lr{i+k}!\lr{j+k}!\lr{i+l}!\lr{j+l}!}\nnl
&~~~~~~~~~~~~~~~~~~~~~~~~\times \out{i+k,j+k}{i+l,j+l} \ .
\end{align*}
In order to calculate the nagativity, we need to partial transpose the matrix above, which can be done very easily
\begin{align*}
\r^{pT}&=\mathcal{N}\sumlim{ijkl=0}{\infty}\frac{g_{+}^{i}g_{-}^{j}z^{k}\bar{z}^l}{i!j!k!l!} \sqrt{\lr{i+k}!\lr{j+k}!\lr{i+l}!\lr{j+l}!}\nnl
&~~~~~~~~~~~~~~~~~~~~~~~~\times \out{i+k,j+l}{i+l,j+k}\\
&=\mathcal{N}\sumlim{ijkl=0}{\infty}\frac{g_{+}^{i}g_{-}^{j}z^{k}\bar{z}^l}{i!j!k!l!} \lr{\ahd}^{i+k}\lr{\bhd}^{j+l}\out{00}{00} {\ah}^{i+l}{\bh}^{j+k} \ .
\end{align*} 
Next we use the relationship and define a shorthand $\r^{pT}\equiv \mathcal{X}$ and the relationships
\begin{align*}
\det\lrb{\exp\lr{A}}=\exp\lr{\Tr{A}}~\Rightarrow~
\det\lrb{X}=\exp\lr{\Tr{\log X}}
\end{align*}
to  derive the characteristic equation and determine the eigenvalues we use
\begin{align*}
\det\lrb{\mathcal{X} -I \lambda}&=\det\lrb{-\lambda I}\det\lrb{I-\frac{\mathcal{X}}{\lambda}}\\
&=\det\lrb{-\lambda I}\exp\lr{\Tr{\log \lr{I-\frac{\mathcal{X}}{\lambda}}}}\\
&=\det\lrb{-\lambda I}\exp\lr{\Tr{ \sumlim{j=1}{\infty}-\frac{\mathcal{X}^j}{j \lambda^j}}}\\
&=\det\lrb{-\lambda I}\exp\lr{ \sumlim{j=1}{\infty}-\frac{\Tr{\mathcal{X}^j}}{j\lambda^j}} \ ,
\end{align*}
where the $\log$ Taylor expansion holds if the eigenvalues of $X$  observe the condition $\lra{\lambda_i}\leq 1$, which is the case for the eigenvalues of any  (partial transposed) density operator.  If we find a general form of $\Tr{X^j}$, then we can hope to find the form of this infinitely long polynomial. Already when calculating a square of $\mathcal{X}$ we can see a pattern,
\begin{align*}
\mathcal{X}^2&=\mathcal{N}^2\sumlim{ijkl=0}{\infty}\sumlim{pqrs=0}{\infty}\frac{g_{+}^{i}g_{-}^{j}z^{k}\bar{z}^l}{i!j!k!l!} \lr{\ahd}^{i+k}\lr{\bhd}^{j+l}\out{00}{00} \\
&\times{\ah}^{i+l}{\bh}^{j+k} \lr{\ahd}^{p+s}\lr{\bhd}^{q+r}\out{00}{00} {\ah}^{p+r}{\bh}^{q+s} \frac{g_{+}^{p}g_{-}^{q}z^{s}\bar{z}^r }{p!q!r!s!} \ .
\end{align*}
Then, focusing on the ket-operator sandwich in the middle we see that
\begin{align*}
\expecfull{00}{ {\ah}^{w}{\bh}^{x} \lr{\ahd}^{y}\lr{\bhd}^{z}}{00}=\delta_{w,y}\delta_{x,z}w!x! \ ,
\end{align*}
in our case implying $p=i+l-s$ and $q=j+k-r$, implying that $i+l\geq s$ and that $j+k\geq r$
\begin{widetext}
{\small \begin{align*}
\mathcal{X}^2&=\mathcal{N}^2\sumlim{ijkl=0}{\infty}\sumlim{r=0}{j+k}\sumlim{s=0}{i+l}\frac{g_{+}^{i}g_{-}^{j}z^{k}\bar{z}^l}{i!j!k!l!} \lr{\ahd}^{i+k}\lr{\bhd}^{j+l}\out{00}{00}  g_+^{i+l-s}g_-^{j+k-r}z^{s}\bar{z}^r\frac{\lr{i+l}!\lr{j+k}!}{\lr{i+l-s}!\lr{j+k-r}!r!s!} {\ah}^{i+l-s+r}{\bh}^{j+k-r+s}\\
&=\mathcal{N}^2\sumlim{ijkl=0}{\infty} \frac{g_{+}^{i}g_{-}^{j}z^{k}\bar{z}^l}{i!j!k!l!} \lr{\ahd}^{i+k}\lr{\bhd}^{j+l}\out{00}{00} \sumlim{r=0}{j+k}\binomial{j+k}{r} g_-^{j+k-r}\bar{z}^{r}{\bh}^{j+k-r}{\ah}^{r}\sumlim{s=0}{i+l} \binomial{i+l}{s}g_+^{i+l-s}z^{s}   {\ah}^{i+l-s}{\bh}^{s}\\
&=\mathcal{N}^2\sumlim{ijkl=0}{\infty}\frac{g_{+}^{i}g_{-}^{j}z^{k}\bar{z}^l}{i!j!k!l!} \lr{\ahd}^{i+k}\lr{\bhd}^{j+l}\out{00}{00}\lr{g_-\bh+\bar{z} {\ah}}^{j+k}\lr{ g_+\ah +z \bh}^{i+l}\\
&=\mathcal{N}^2\sumlim{i=0}{\infty}\frac{g_+^{i}}{i!} \lrb{\ahd \cdot\lr{ g_+\ah +z \bh}}^{i}
\sumlim{j=0}{\infty}\frac{g_-^{j}}{j!} \lrb{\bhd\cdot\lr{ \bar{z}\ah +g_- \bh}}^{j}
\sumlim{k=0}{\infty}\frac{z^{k}}{k!} \lrb{\ahd\cdot\lr{ \bar{z}\ah +g_- \bh}}^{k}
\sumlim{l=0}{\infty}\frac{\bar{z}^{l}}{l!} \lrb{\bhd\cdot\lr{ g_+\ah +z \bh}}^{l}\out{00}{00} \\
&=\mathcal{N}^2 \exp\lrb{g_+ \ahd\cdot\lr{ g_+\ah +z \bh}+g_-   \bhd \cdot\lr{ \bar{z}\ah +g_- \bh}+z \ahd \cdot\lr{ \bar{z}\ah +g_- \bh}+\bar{z} \bhd \cdot\lr{ g_+\ah +z \bh}}\\
&=\mathcal{N}^2 \exp\lrb{\lr{f_3^2+f_5^2+f_9^2+f_{12}^2}H_3+2f_3f_9 H_9+2f_3\sqrt{f_5^2+f_{12}^2}\lr{\sin\lr{\theta+\varphi} H_5^{pT}+\cos\lr{\theta+\varphi} H_{12}^{pT}}}\out{00}{00} \ .
\end{align*} }
\end{widetext}
Since the trace is unaffected by (partial) transposition, and the trace of $\mathcal{X}^2/\mathcal{N}^2$ is the same as the trace of $\r/\mathcal{N}$ with the replacement of $f_3\to f_3^2+f_5^2+f_9^2+f_{12}^2$, $f_9\to 2f_3f_9$ and $f_5^2+f_{12}^2\to 4f_3^2\lr{f_5^2+f_{12}^2}$, giving after simplification
\begin{align*}
\Tr{\mathcal{X}^2}=\frac{\lr{1-x_+^2}\lr{1-x_-}^2}{\lr{1-x_+^2}\lr{1-x_-^2}} \,.
\end{align*}
Following the argument above it is easy to prove in general (after some algebra)  that every additional power of $\mathcal{X}$ gives rise to the transformation $\lr{\cdot}\ah\to  g_+\lr{\cdot}\ah +z \lr{\cdot}\bh$ and $\bh\to \bar{z}\lr{\cdot}\ah +g_- \lr{\cdot}\bh$. Using a proof by induction one can prove that upon tracing $\mathcal{X}^j$ we get
\bea
\Tr{\mathcal{X}^j}=\frac{\lr{1-x_+}^j\lr{1-x_-}^j}{\lr{1-x_+^j}\lr{1-x_-^j}} \,. \label{XjTrace}
\eea
Alternatively, one can see that $\mathcal{X}^j$ can always be written in the form
{\footnotesize \bea
\mathcal{X}^j&=&\mathcal{N}^j \exp\lrb{F_3^{\lr{j}} H_3+F_9^{\lr{j}} H_9+F_5^{\lr{j}} H_5^{pT}+F_{12}^{\lr{j}}H_{12}^{pT}}\,,\label{XjAnsatz}
\eea}
with $\Tr{\mathcal{X}^j}$ in the form
\bea
\Tr{\mathcal{X}^j}=\frac{\mathcal{N}^j}{\lr{1-X_+^{\lr{j}}}\lr{1-X_-^{\lr{j}}}}\,,\label{TraceAnsatz}
\eea
with
\bea
X_{\pm}^{\lr{j}}=F^{\lr{j}}_3\pm\sqrt{\lr{F^{\lr{j}}_5}^2+\lr{F^{\lr{j}}_9}^2+\lr{F^{\lr{j}}_{12}}^2}\,.~~~~
\eea
By multiplying both sides of equation (\ref{XjAnsatz}) by $\mathcal{X}$ one can arrive at a set of recursive linear algebraic equations for functions $F_i^{\lr{j}}$
\begin{align*}
F_3^{j+1}&=f_{12}F_{12}^{j}+f_{3}F_3^{j}+f_{5}F_5^{j}+f_{9}F_{9}^{j}\,, \\
F_{9}^{j+1}&=-if_{12}F_5^{j}+if_{5}F_{12}^{j}+f_{3}F_{9}^{j}+f_{9}F_3^{j}\,, \\
F_{5}^{j+1}&=if_{12}F_{9}^{j}-if_{9}F_{12}^{j}+f_{3}F_5^{j}+f_{5}F_3^{j}\,, \\
F_{12}^{j+1}&=f_{12}F_3^{j}+f_{3}F_{12}^{j}+i(f_{9}F_5^{j}-f_{5}F_{9}^{j})\,, \\
\end{align*}
  with solutions
\begin{align*}
F_3^{j}&=\frac{1}{2} \left(x_+^{j}+x_-^{j}\right)\,, & F_{9}^{j}&=\frac{f_{9} \left(x_+^{j}-x_-^{j}\right)}{x-y}\,, \\
F_5^{j}&=\frac{f_{5} \left(x_+^{j}-x_-^{j}\right)}{x-y}\,, &  F_{12}^{j}&=\frac{f_{12} \left(x_+^{j}-x_-^{j}\right)}{x-y}\,, 
\end{align*}
which when substituted into the equation (\ref{TraceAnsatz}) yield again equation (\ref{XjTrace}).

We will now use this result to calculate   the eigenvalues of $\mathcal{X}=\r^{rm pTr}$ as
\begin{align*}
\det\lrb{\mathcal{X}-I \lambda}&={\det\lrb{-\lambda I}} \exp\lr{ \sumlim{j=1}{\infty}-\frac{\Tr{\mathcal{X}^j}}{j\lambda^j}}\\
&={\det\lrb{-\lambda I}} \exp\lr{ -\sumlim{j=1}{\infty}\frac{\lr{1-x}^j\lr{1-y}^j}{j\lambda^j\lr{1-x^j}\lr{1-y^j}}} \ .
\end{align*}
Let us define
$$
\sigma=\frac{\lambda}{\mathcal{N}}=\frac{\lambda}{\lr{1-x_+}\lr{1-x_-}} \ ,
$$
then this in combination with
$$
\frac{1}{1-r}=\sumlim{i=0}{\infty} r^i
$$
gives us
\begin{align*}
\det\lrb{\mathcal{X}-I \lambda}&={\det\lrb{-\lambda I}} \exp\lr{ -\sumlim{j=1}{\infty}\frac{1}{j\sigma^j}\sumlim{p=0}{\infty} x_+^{pj}\sumlim{q=0}{\infty} x_-^{jq}}\\
&={\det\lrb{-\lambda I}} \exp\lr{ \sumlim{p,q=0}{\infty}\sumlim{j=1}{\infty}-\frac{1}{j} \lr{\frac{x_+^{p} x_-^{q}}{\sigma}}^j}\\
&={\det\lrb{-\lambda I}} \exp\lr{ \sumlim{p,q=0}{\infty}\log\lr{1-\frac{x_+^{p} x_-^{q}}{\sigma}}}\\
&={\det\lrb{-\lambda I}} \prod\limits_{p,q=0}^{\infty}\lr{1-\frac{x_+^{p} x_-^{q}\mathcal{N}}{\lambda}} \ ,
\end{align*}
so that when the above is equal to zero, it is easy to see that all of the eigenvalues $\lambda_i$ are of the form $x_+^{p} x_-^{q} \mathcal{N}$. Since $x_+>0$ and $x_-<0$ if $\sqrt{f_5^2+f_9^2+f_{12}^2}>f_3$, then the only negative eigenvalues will be present for odd powers of $x_-$ and any power of $x_+$. Thus negativity takes the form given by equation (\ref{eq:negativity}).

 \subsection{The smallest non-trivial problem, the largest with analytically obtainable transient solutions}
 The master equation (\ref{masterEq}) can be solved exactly for an arbitrary initial condition under the parameter reduction \textbf{5}, i.e. $\k_1=\k_2=\k$ and $n_{{\rm th},1}=n_{{\rm th},2}=n_{{\rm th}}$,
using the normal ordering solution Ansatz
 \begin{align*}
\r\lr{t}&=e^{f_0\lr{t}} e^{f_3\lr{t}H_3}e^{f_5\lr{t}H_5}e^{f_{1}\lr{t}H_{1}}e^{f_6\lr{t}H_6}\\
&\hspace{4cm}\times e^{f_2\lr{t}H_2}e^{f_4\lr{t}H_4}\r\lr{0}\,,\nn
 \end{align*}
and the Wei-Norman method \cite{WeiNorman63,WeiNorman64}  we obtain equations

\begin{align*}
\dot{f}_1&= \frac{1}{2} (\k  (2 (n_{\rm th}+1) f_3 -2 n_{\rm th}-1)-\frac{1}{2}\lra{\xi} f_5 ) \\
\dot{f}_2&= \frac{1}{2} e^{2 f_1 } (2 \k  (n_{\rm th}+1) \cosh f_6 -\frac{1}{2}\lra{\xi} \sinh f_6 ) \\
\dot{f}_3&= - \frac{1}{2} \lra{\xi} f_3 f_5 -\k\lr{ 2  n_{\rm th}+1}f_3+\k  (n_{\rm th}+1) f_3 ^2\\
&\hspace{2cm}+\k  \left((n_{\rm th}+1) f_5 ^2+n_{\rm th}\right) \\
\dot{f}_4&= \frac{1}{2} e^{2 f_1 }\lr{ \frac{1}{2} \lra{\xi} \cosh f_6 -2 \k  (n_{\rm th}+1) \sinh f_6 } \\
\dot{f}_5&=    \k  f_5  (2 (n_{\rm th}+1) f_3 -2 n_{\rm th}-1)-\frac{1}{4}\lra{\xi} \lr{f_3 ^2+f_5 ^2-1}  \\
\dot{f}_6&= 2 \k  (n_{\rm th}+1) f_5 -\frac{1}{2}\lra{\xi} f_3
\end{align*}
and the last one being $\dot{f}_{0}\lr{t}=\k$. These equations linearly decompose into two sets of equations for functions $\left\{p_+,q_+,r_+\right\}$ and $\left\{p_-,q_-,r_-\right\}$ such that
\begin{align*}
f_{1}\lr{t}&= \frac{1}{4} (p_{-}\lr{t}+p_{+}\lr{t}) &   f_{4}\lr{t}&= \frac{1}{2} (q_{+}\lr{t}-q_{-}\lr{t}) \\
f_{2}\lr{t}&= \frac{1}{2} (q_{-}\lr{t}+q_{+}\lr{t}) &   f_{5}\lr{t}&= \frac{1}{2} (r_{-}\lr{t}-r_{+}\lr{t})\\
f_{3}\lr{t}&= \frac{1}{2} (r_{-}\lr{t}+r_{+}\lr{t}) &   f_{6}\lr{t}&= \frac{1}{2} (p_{-}\lr{t}-p_{+}\lr{t})
\end{align*}
where
\begin{align*}
p_{\pm}\lr{t}&=2 \log \left(\frac{2 e^{\frac{1}{2} t \lr{\k \pm\lra{\xi}/2 }} (\k \pm\lra{\xi}/2 )}{-2 n_{{\rm th}} \k +e^{t \lr{\k \pm\lra{\xi}/2 }} \lr{2 \lr{n_{{\rm th}}+1} \k \pm\lra{\xi}/2}\pm\lra{\xi}}\right) \ , \\
 q_{\pm}\lr{t}&=\frac{\left(1-e^{t \lr{\k \pm\lra{\xi}/2 }}\right) \lr{2 \lr{n_{{\rm th}}+1} \k \pm\lra{\xi}/2}}{\left(1-e^{t \lr{\k \pm\lra{\xi}/2 }}\right) \lr{2 \lr{n_{{\rm th}}+1} \k \pm\lra{\xi}/2}-2 \lr{\k \pm\lra{\xi}/2 }} \ , \\
 r_{\pm}\lr{t}&=\frac{\left(1-e^{t \lr{\k \pm\lra{\xi}/2 }}\right) \lr{\pm\lra{\xi}/2-2 n_{{\rm th}} \k }}{-2 n_{{\rm th}} \k +e^{t \lr{\k \pm\lra{\xi}/2 }} \lr{2 \lr{n_{{\rm th}}+1} \k \pm\lra{\xi}/2}\pm\lra{\xi}/2} \ .
 \end{align*}

\subsection{Details of the initial thermal state computation}
If the system is initially in the thermal state $\r\lr{0}=\lr{1-e^{-\beta \hbar \omega }}\exp\lr{-\b\hbar \o\ahd\ah}=\lr{1-e^{-\beta \hbar \omega }}\exp\lr{e^{-\b\hbar \o} H_3}\out{0}{0}\equiv\lr{1-\tau}e^{\tau H_3}\out{0}{0}$, then the evolution takes the form
\begin{align*}
\r &=\lr{1-\tau}^2 e^{f_0\lr{t}} e^{f_3\lr{t}H_3}e^{f_5\lr{t}H_5}e^{f_1\lr{t}H_1}e^{f_6\lr{t}H_6}\\
&\hspace{3.2cm}\times e^{f_2\lr{t}H_2}e^{f_4\lr{t}H_4}e^{\tau H_3}\out{0}{0}\,.\nn
\end{align*}
This can be rewritten again in the form involving only exponents of $H_3$ and $\mathscr{H}_{\theta,5}$, by means of sandwiching the last exponent in the following manner
\begin{align*}
\r&=\lr{1-\tau}^2 e^{f_0 } e^{f_3 H_3}e^{f_5 H_{5}}e^{f_1 H_1}e^{f_6 H_{6}}e^{f_2 H_2}e^{f_4 H_{4}}e^{\tau H_3}\\
&~~~\times e^{-f_4 H_{4}}
 e^{-f_2 H_2}e^{-f_6 H_{6}}e^{-f_1 H_1} e^{f_1 H_1}e^{f_6 H_{6}}e^{f_2 H_2}e^{f_4 H_{4}}\out{0}{0}\nnl
&=\lr{1-\tau}^2 e^{f_0 +\frac{1}{2}f_1 } e^{f_3 H_3}e^{f_5 H_{5}}e^{f_1 H_1}e^{f_6 H_{6}}e^{f_2 H_2}e^{f_4 H_{4}}e^{\tau H_3}\nnl
&~~~~~~\times e^{-f_4 H_{4}} e^{-f_2 H_2}e^{-f_6 H_{6}}e^{-f_1 H_1} \out{0}{0}\,,\nn
\end{align*}
which can be brought back to an easier form by realising that
\bea
e^{f_1 H_1}e^{f_6 H_{6}}e^{f_2 H_2}e^{f_4 H_{4}}  H_3 e^{-f_4 H_{4}}e^{-f_2 H_2}e^{-f_6 H_{6}}e^{-f_1 H_1}\nnl
= \sumlim{i=1}{6}A_i H_i\equiv  \mathcal{J}\,,\nn
\eea
with
\begin{align*}
 A_1=2 f_2\,,  & \hspace{2cm} A_3=e^{f_1} \cosh  f_6 \,, \\
 A_6=2 f_4\,,  & \hspace{2cm} A_5=e^{f_1} \sinh f_6 \,,
\end{align*}
and
\begin{align*}
A_{2,4}&=\frac{1}{2}\lr{e^{-\lr{2f_1+f_6}}\lr{f_2+f_4}^2\pm e^{-\lr{2f_1-f_6}}\lr{f_2-f_4}^2}\,,
\end{align*}
and that $e^{A}e^{\tau B} e^{-B}=e^{\tau e^A B e^{-A}}$, gives
\begin{align*}
&= e^{f_1 H_1}e^{f_6 H_{6}}e^{f_2 H_2}e^{f_4 H_{4}} e^{\tau H_3} e^{-f_4 H_{4}}e^{-f_2 H_2}e^{-f_6 H_{6}}e^{-f_1 H_1}\nnl
&= \exp\lrb{\tau \sumlim{i=1}{6}A_i H_i}=e^{\mathcal{\tau J}}
\end{align*}
and we set out to find $\mathcal{F}_i$ such that
\begin{align*}
e^{\mathcal{\tau J}}=e^{\mathcal{F}_3 H_3}e^{\mathcal{F}_5 H_{5}}e^{\mathcal{F}_1 H_1}e^{\mathcal{F}_6 H_{6}}
e^{\mathcal{F}_2 H_2}e^{\mathcal{F}_4 H_{4}}
\end{align*}
is another  normal ordering  decomposition Ansatz of an operator exponent. This time however it is not a decomposition based on time evolution, but rather the initial condition parameter $\tau$ is acting like  an artificial evolution operator which ranges from 0 ($k_bT\ll\hbar \o$) to 1 ($k_bT\gg\hbar \o$).
We derive a set of differential equations  for functions $\mathcal{F}_i$ (the Wei-Norman method) based on
\begin{align*}
\partial_{\tau}e^{\tau \mathcal{J}}
=\mathcal{J}e^{\mathcal{F}_3H_3}e^{\mathcal{F}_5H_{5}}e^{\mathcal{F}_1H_1}e^{\mathcal{F}_6H_{6}}e^{\mathcal{F}_2H_2}e^{\mathcal{F}_4H_{4}}
\,,
\end{align*}
with the solutions
\begin{align*}
\mathcal{P}_{\pm}&= 2 \log (1-\tau  q_{\pm}) &
\mathcal{Q}_{\pm}&= \frac{\tau  e^{-p_{\pm}} q_{\pm}^2}{1-\tau  q_{\pm}} \\
\mathcal{R}_{\pm}&= \frac{\tau  e^{p_{\pm}}}{1-\tau  q_{\pm}}
\end{align*}
with
\begin{align*}
\mathcal{F}_{1}\lr{t}&= \frac{1}{4} (\mathcal{P}_{-}\lr{t}+\mathcal{P}_{+}\lr{t}) &   \mathcal{F}_{4}\lr{t}&= \frac{1}{2} (\mathcal{Q}_{+}\lr{t}-\mathcal{Q}_{-}\lr{t}) \\
\mathcal{F}_{2}\lr{t}&= \frac{1}{2} (\mathcal{Q}_{-}\lr{t}+\mathcal{Q}_{+}\lr{t}) &   \mathcal{F}_{5}\lr{t}&= \frac{1}{2} (\mathcal{R}_{-}\lr{t}-\mathcal{R}_{+}\lr{t})\\
\mathcal{F}_{3}\lr{t}&= \frac{1}{2} (\mathcal{R}_{-}\lr{t}+\mathcal{R}_{+}\lr{t}) &   \mathcal{F}_{6}\lr{t}&= \frac{1}{2} (\mathcal{P}_{-}\lr{t}-\mathcal{P}_{+}\lr{t})
\end{align*}
and then the density operator reads
\begin{align*}
\r&=\lr{1-\tau}^2 e^{f_0 +\frac{1}{2}f_1 } e^{f_3 H_3}e^{f_5 H_{5}}e^{\mathcal{F}_3 H_3}e^{\mathcal{F}_5 H_{5}}e^{\mathcal{F}_1 H_1}
e^{\mathcal{F}_6H_{6}}\nnl
&\hspace{3cm}\times e^{\mathcal{F}_2 H_2}e^{\mathcal{F}_4H_{4}}\out{0}{0}\,,\nn
\end{align*}
which upon the action of the annihilation operators on the vacuum state yields
\bea
\r &=&\lrb{\lr{1- g_3\lr{t} }^2- g_5\lr{t}  ^2}^{1/2}e^{ g_3\lr{t} H_3}e^{g_5H_{5}}\out{0}{0}\,,\nn
\eea
where $g_i\lr{t}=f_i\lr{t}+\mathcal{F}_i\lr{t}$.

It is important to note that only functions $\mathcal{F}_i$ carry the information about the initial thermal state stored in the variable $\tau$, and that in the final result only $\mathcal{F}_{3}$ and $\mathcal{F}_{5}$ remain relevant. What is very interesting is that these two functions in the steady state vanish, i.e. $\lim\limits_{t\to\infty} \mathcal{F}_{3}=0=\lim\limits_{t\to\infty} \mathcal{F}_{5}$. This means that any impact of this initial state parameter $\tau$ is completely irrelevant to the steady state entanglement of the system on both  sides the parameter regimes boundary $\lra{\xi}=2\k$.

\end{document}